\documentstyle[aps,preprint,epsf]{revtex}
\begin{document}
\thispagestyle{empty}
\begin{center}
{\large \bf
Solar Proton Burning Process Revisited within 
a Covariant \\
Model Based on the Bethe-Salpeter Formalism }\\[1cm]
{\sc
L.P. Kaptari$^{a,b}$,
B. K\"ampfer$^{a},$
E. Grosse$^{a}$
}\\[1cm]

$^a$Forschungszentrum Rossendorf,
PF 510119, 01314 Dresden, Germany\\[1mm]
$^b$Bogoliubov Laboratory of Theoretical Physics, JINR Dubna,\\
141980, Moscow reg., Russia \\[1mm]
\end{center}
%
\begin{abstract}
A covariant model based on the Bethe-Salpeter formalism is proposed
for investigating the solar proton burning process
$pp\to De^+\nu_e$ and the near-threshold deuteron disintegration via 
electromagnetic and weak interactions. 
Results of numerical calculations of the energy dependence of relevant cross
sections and the astrophysical low-energy cross section factor
$S_{pp}$ of the proton burning process are presented.
Our results confirm previous canonical values, and the energy dependence of 
the $S_{pp}$ factor is rather close to phenomenological extrapolations 
commonly  adopted  in computations of solar nuclear reaction rates. 
\end{abstract}
\newpage
\setcounter{page}{1}
{\bf I. Introduction:}
The proton-proton fusion reaction
$pp\to De^+\nu_e$ plays an important role in understanding processes
occurring in the Sun and in investigating the solar structure. 
This reaction is the first step in the chain of nuclear reactions 
producing the solar energy and neutrino
fluxes. Unfortunately the corresponding 
cross section is far too low for a direct laboratory study.
In this context reliable theoretical calculations
and estimates of the reaction rates are of a great importance. Nowadays, the
most precise estimates have been obtained within the framework
of low-energy potential models \cite{bahcall,adelb,riska,panda}. 
The accuracy of the obtained results is believed to be
of the order of few per cents or even better \cite{adelb,kamion}. 
For instance, 
the low-energy cross section factor $S_{pp}$ that determines the  
rates for the proton-proton fusion 
is estimated as 
$S_{pp} = 4.00 (1 \pm 0.007) \times 10^{-25}$ MeV b \cite{adelb}. 
While this number is considered as a canonical value, recent
estimates \cite{ivanov},  based on an 
effective relativistic field theory,
claimed a considerably larger value being inconsistent with helioseismic
data \cite{astro133,critica}. 
This has triggered a series of recalculations of $S_{pp}$
within various approaches (cf.\ \cite{nucl0021,nucl0066,astro144} and
further references therein), which recover values of $S_{pp}$
centered around the canonical one.
However, there are some other indications \cite{alfio}
that $S_{pp}$ may be $4.2 \times 10^{-25}$ MeV b or even larger.
 
Calculations of the $pp$ rates require the evaluation of  two
main ingredients: (i) the weak-interaction matrix element, 
and (ii) the overlap
of the $pp$ and deuteron wave functions. In most relativistic 
effective field theories a self-consistent treatment of the deuteron bound
state is difficult, so one usually adopts the short-range effective theory 
\cite{wise}
to avoid an explicit use of the deuteron wave function. Within
such models one may achieve  good fits of data in the zero energy limit,
whereas  a computation of matrix elements at finite values of the energy 
is not yet available. 
 
In this paper we  calculate the relevant matrix elements within
a fully covariant model of the deuteron,
based on the Bethe-Salpeter (BS) formalism.  Within this model 
a variety of quantities, including the total and differential
cross sections, energy dependences, angular distributions etc.\ 
are accessible.
In the first part of this paper we introduce our formalism and
consider as examples the 
processes of near-threshold deuteron disintegration
by electromagnetic and weak interactions. 
The second part extends the formalism and 
is entirely dedicated to the solar $pp$ burning processes. Within
the relativistic impulse approximation a fully covariant expression for 
the $S_{pp}$ factor is obtained.

{\bf II. Deuteron disintegration near thresholds:}
In this section we consider processes of the type
$ l D \to l' N_1 N_2$, i.e. the 
deuteron disintegration  by leptons in electromagnetic
($eD\to e'pn$) and weak ($e^+D\to\nu pp$) reactions near threshold.
There the amplitudes 
fall down  rapidly and the cross sections are governed by the phase space of
final products. The threshold energies
are  $E_{thr}=2.7375$ MeV ($eD$ reactions)
and  $E_{thr}= 0.93139$ MeV (weak interaction), 
where $E_{thr}$ is the total energy
of the incident lepton in the deuteron  rest frame. 
We consider the electromagnetic disintegration process within the
one-photon exchange approximation (cf. fig.~1a) and the weak 
positron-nucleon interaction (cf.\ fig.~1b)
within the effective Fermi model. 
The deuteron vertex $D\to NN$ is treated 
in a covariant way within the  BS formalism.
The  BS equation has been solved numerically \cite{solution},
with  realistic one-boson-exchange interactions, and suitably
parameterized \cite{parametrization}.
The  solution describes fairly accurate the main low-energy and 
static characteristics of the deuteron, such as the deuteron binding energy, 
and its quadrupole and magnetic moments \cite{quad}.

\begin{figure}[ht]
\epsfbox{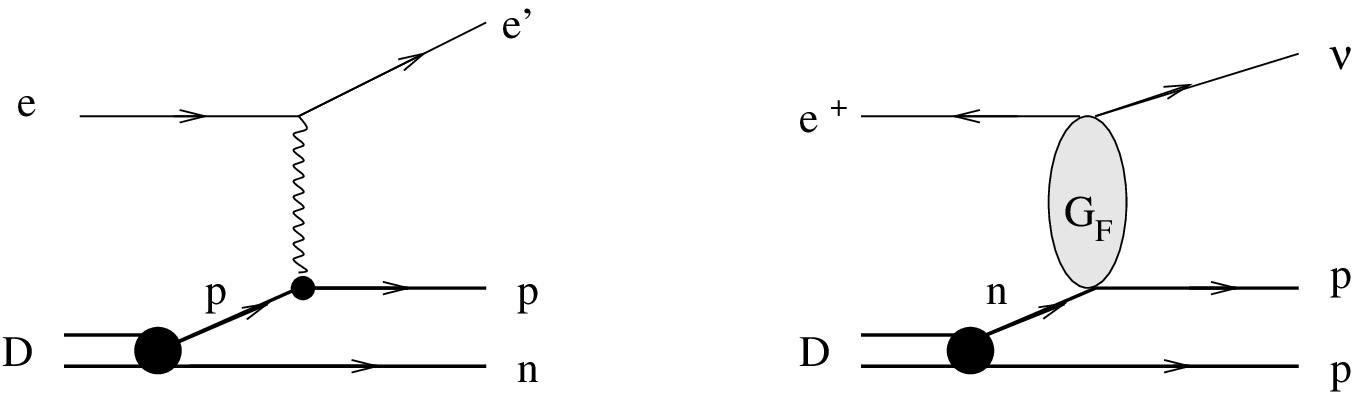}
\vskip 1cm
\caption{ Feynman diagrams for electromagnetic (left panel)
and weak (right panel) deuteron disintegration. The weak coupling constant 
and the propagator of the $W^+$ boson
has been replaced by the effective Fermi constant $G_F$ for the 4-point contact interaction.}
\label{fig1}
\end{figure}

The invariant
cross section of the process $D(l,l')NN$, depicted in fig.~\ref{fig1}, 
has the form
\begin{eqnarray}
d\sigma = \frac{(2\pi)^4\delta^{(4)}\left( k_l+p_D -k_l'- p_{N_1}-p_{N_2} 
\right) }
{2\sqrt{\lambda(s,M_d^2,\mu^2)}}\left|  {\cal M}_{lD\to l' NN }\right |^2
\frac{ d{\bf  k_l'}}{(2\pi)^3 2{\cal E}'}  
\frac{d{\bf  p_1'}}{(2\pi)^3 2 E_1'}\frac{d{\bf  p_2'}}{(2\pi)^3 2 E_2'},
\label{cs00}
\end{eqnarray}
where $\lambda(s,M_d^2,\mu^2)\equiv[s-(M_d+\mu)^2][s-(M_d-\mu)^2]$ 
is a kinematical factor with $s$ as usual Mandelstam variable,
$M_d$ and $\mu$  are  the deuteron and the electron (positron) 
masses, respectively. The 4-momenta
of leptons are denoted as $ k=({\cal E},{\bf k})$, 
while for hadron momenta the notation $ p=( E,{\bf p})$ is used. The prime superscript
specifies the momenta of  the final particles.
The subscript ``2''  denotes the spectator nucleon, i.e.
the neutron in $eD$ reactions and the non-interacting
proton in $e^+D$ processes (see fig.~\ref{fig1}).
The two nucleons in the final state are treated in the plane wave 
approximation,
i.e. the effects of final state interaction are here disregarded and all 
calculations are performed within the relativistic impulse approximation.

The square of the invariant amplitude
${\cal M}_{lD\to l' NN }$ may be cast in the form
$
|{\cal M}_{l D\to l'  N N }|^2=S^2(Q^2)\,l_{\mu\mu}\, W^{\mu\nu},
$
where $S(Q^2=-(k_l-k_l')^2)$  
is the Feynman propagator of the intermediate exchanged
boson (photon or $W^+$ boson), and the leptonic ($l_{\mu\nu}$) and hadronic
($W_{\mu\nu}$) tensors are defined by
\begin{eqnarray}
&&
l_{\mu\nu} =\frac12 \sum\limits_{\rm spins} \langle l' |\hat j_\mu| l\rangle\langle l |\hat j_\nu^+| l'\rangle
=\frac12\rm {Tr}\left[ (\hat k_l+\mu)\Gamma_\mu (\hat k_l'+\mu')\Gamma_\nu\right],
\label{lmunu}
\\&&      
W_{\mu\nu} =\frac13\sum\limits_{\rm spins}\langle N_1'N_2' |\hat J_\mu|D\rangle\langle D |\hat J_\nu^+|
N_1'N_2'\rangle,
\label{dtensor}
\end{eqnarray}
where $\hat j_\mu$ and $\hat J_\mu$ denote the leptonic and hadronic current 
operators, $\Gamma_\mu$ is the corresponding 
vertex function, related to the choice of interacting  currents;
$\hat k$ is used as a short hand notation for $\gamma^\mu k_\mu$.
The electron-nucleon {\em electromagnetic} vertex in the on-mass-shell form, 
$
\Gamma^{eN}_{\mu}(Q^2) = \gamma_{\mu} F_1(Q^2) + 
i \frac{\sigma_{\mu\alpha}Q^{\alpha}}{2m}\kappa F_2(Q^2),
$
contains  the 
electromagnetic form factors of the nucleon  $F_{1,2}$, and   is its 
anomalous magnetic moment $\kappa$.
In view of the small transferred momenta and low initial energies, 
considered here,
the second term 
may be safely disregarded, and with high accuracy also the nucleon formfactor
$F_1(Q^2)$ is replaced by unity. 
Then the electromagnetic part of the leptonic tensor
is computed straightforwardly. 
For the {\em weak vertices} one may employ the Weinberg-Salam
theory for weak interactions, however,  
in the considered energy range the effective
Fermi model is appropriate. Thus
\begin{eqnarray}
&&
\hat j_\mu(x) = \bar\psi_e(x)\gamma_\mu\, (1-\gamma_5)\psi_\nu(x),\label{tokl}\\
&&
\hat J_\nu(x)^{(\Delta S=0)} = \cos\theta_c
\bar\psi_N(x)
\frac{(\tau_1-i\tau_2)}{\sqrt{2}}
\gamma_\mu\, (1-R\gamma_5)\psi_N(x),\label{tokD}
\end{eqnarray}
and $S(Q^2) = \displaystyle\frac{G_F}{\sqrt{2}}$, and
$R=1.2681$  is the ratio of axial to vector coupling constants,
$G_F=1.16639\times 10^{-5}\,\rm GeV^{-2}$ is the effective Fermi constant
for the weak nucleon decay, $\theta_c=0.24$ is the Cabibbo strangeness
 mixing angle ($\Delta S=0$ means the strangeness
conserving part of the current),
$\psi_{l,N}(x)$ are the corresponding leptonic and nucleonic field
operators, $\tau$  stands for the isospin Pauli  matrices. 

With virtue of the Mandelstam technique\cite{mand} for computing
matrix elements within the BS formalism, the deuteron tensor $W_{\mu\nu}$ in 
eq.~(\ref{dtensor})
receives the form
\begin{eqnarray}
&&
W_{\mu\nu}=\frac13\,(p_2^{\,\prime \, 2}-m_2^2){\rm Tr}\left [
\bar\Psi^{BS}(p_1,p_2)\Gamma_\mu\,(\hat p_1^{\,\prime}+m_1)\,\Gamma_\nu\,
\Psi^{BS}(p_1,p_2)\,(\hat p_2^{\, \prime}+m_2)\right ],
\label{tr}
\end{eqnarray}
where $\Psi^{BS}$ is the deuteron amplitude as defined in 
refs.~\cite{quad,cubis,keisterTj} and
$p_1$ and $p_2=p_2'$ are the momenta of the deuteron constituents 
with a mass $m$, i.e.
the proton ($m_p$) or neutron ($m_n$) masses.
By inspecting  eq.~(\ref{tr}) one  observes that,
due to the adopted mechanism in fig.~\ref{fig1}, the spectator 
nucleon is on mass shell,  and the first
factor in eq.~(\ref{tr}) is always zero. Nevertheless,  the BS amplitudes
are themselves singular when one nucleon is on mass shell 
with the singularity of each
amplitude being of the order $(p_2^2-m_2^2)^{-1}$. Consequently, the 
r.h.s.\ of  eq.~(\ref{tr}) is finite.
In the general case the BS amplitude  $\Psi^{BS}$, which is a 
$4 \times 4$
matrix, may be projected on   a set of basis matrices and such an operation
determines eight independent  partial amplitudes \cite{quad,cubis}. 
However not all these amplitudes equally contribute
to the deuteron matrix elements. One may use a basis set of
spin-angular matrices \cite{cubis}
which allows to classify the partial amplitudes corresponding to their
contributions in a full   analogy with the non-relativistic case as
$^3S_1^{++(--)}$,  $^3D_1^{++(--)}$, $^3P_1^{+-(-+)}$,   
$^1P_1^{+-(-+)}$  waves in the spectroscopical notation
(the superscripts denote the $\rho$ spin of each constituent, 
cf.\ \cite{quad} for details). 
Then the $^3S_1^{++}$ and  $^3D_1^{++}$  waves 
turn out to be a relativistic generalization
of the non-relativistic $S$ and $D$ waves in the deuteron\cite{quad}. 
In the considered reactions
the contribution of $D$ waves vanishes 
near the threshold (as do the other amplitudes with at least one negative 
$\rho$ spin),
leaving us with  contributions only from the $^3S_1^{++}$ waves. 
A direct evaluation of traces  in eqs.~(\ref{lmunu}, \ref{tr}) results in 
analytical, covariant expressions for the cross section. For instance,
the electromagnetic tensors read 
\begin{eqnarray}
l_{\mu\nu}\, W^{\mu\nu}
=&&16M_d(2\pi)^3\Phi^2(p_0,|{\bf p}_2|)\times\nonumber\\
&&
\left[(k_e p_1')(k'p_1)-\mu^2(p_1p_1')-(k_e'k_e)m_p^2+(k_e'p_1')(k_ep_1)+
2\mu^2 m_p^2
\right],
\label{xsect}
\end{eqnarray}
where $\Phi(p_0,|{\bf p}_2|)$ is the covariant  
generalization of the deuteron wave
function \cite{quad} within the BS formalism, which  has the
following form in the deuteron rest frame
\begin{equation}
\Phi(|{\bf p}_2|)=\frac{G_S(p_0,|{\bf p}_2|)}{(2\pi)^2\sqrt{2M_d}
(M_d-2E_{p})^2}\,\, , \quad\quad p_0=M_d/2-E_{p}, \quad E_p=\sqrt{m^2+{\bf p}_2^2},
\label{psis}
\end{equation}
where $G_S(p_0,|{\bf p}_2|)$ is the BS partial vertex corresponding to the 
$^3S_1^{++}$ amplitude~\cite{quad} and $m$ is an average value of 
the nucleon mass.
Inserting eq.~(\ref{xsect}) into eq.~(\ref{cs00}) yields
\begin{eqnarray}
&&
d\sigma^{eD}=\frac{m_n }{E_2} \,
\Phi^2(|{\bf p}_2|) \, d{\bf p}_2 \,\delta(p_1^{\,'\,2}-m_p^2)
\times \label{ed}\\
&&\!\!\!
\left\{
\frac{4\alpha^2}{Q^4}  
\frac{4M_d}{m_p\sqrt{\lambda(s,M_d^2,\mu^2)} }
\left( 
(k_e p_1')(k'p_1)-\mu^2(p_1p_1')-(k_e'k_e)m_p^2+(k_e'p_1')(k_ep_1)+2\mu^2 m_p^2
\right )
\frac{ d{\bf k}_e'}{2{\cal E}'}
\right\},
\nonumber
\end{eqnarray}
where the expression   enclosed in the curly  brackets 
together with the $\delta$ function  is  merely  the cross section 
$d\sigma^{ep}$ of
an electron scattering off a moving proton with the momentum $p_1$. 
Hence, for the 
electromagnetic $eD$ disintegration processes the cross section  may be 
written  as
\begin{equation}
\sigma^{eD}=\int
 d{\bf p}_2 \,\frac{m_n }{E_2}\, 
\Phi^2(|{\bf p}_2|)\,d\sigma^{ep}(k,p_1).
\label{totale}
\end{equation}
By processing in an analogous way with the weak disintegration diagram 
displayed in fig.~1b, one obtains
\begin{equation}
\sigma^{e^+D}_w=\int
 d{\bf p}_2 \,\frac{m_n }{E_2}\, 
\Phi^2(|{\bf p}_2|)\,d\sigma^{en}_w(k,p_1),
\label{totalw}
\end{equation}
where, for convenience, the ``elementary'' cross section 
$d\sigma^{en}_w(k,p_1)$
for a hypothetical $e^+n\to \nu p$ reaction
has been introduced by
\begin{eqnarray}
d\sigma^{en}_w(k,p_1) & = & \frac{G_F^2}{(2\pi)^2}
\frac{4M_d\delta(p_1^{\,'\,2}-m_p^2)}{m_p\sqrt{\lambda(s,M_d^2,\mu^2)} }
\left\{
\left(R^2+1)\right)
\left[
(k_ep_1)(k_\nu'p_1')+(k_ep_1')(k_\nu'p_1)\right ] \right .\nonumber\\
&& + \left .
\left(R^2-1\right))m_pm_n(k_ek_\nu')-
2R\left[
(k_ep_1')(k_\nu'p_1)-(k_ep_1)(k_\nu'p_1')\right ]\right\}
\frac{ d{\bf k}_\nu'}{{\cal E}'}.
\label{elemw}
\end{eqnarray}
In computing   eq.~(\ref{totalw}) the
angular integration on ${\bf p}_2$ should be performed 
in only one-half of the hemisphere
due to the two identical particles in the final state. 

\begin{figure}
\epsfxsize 5in
\epsfbox{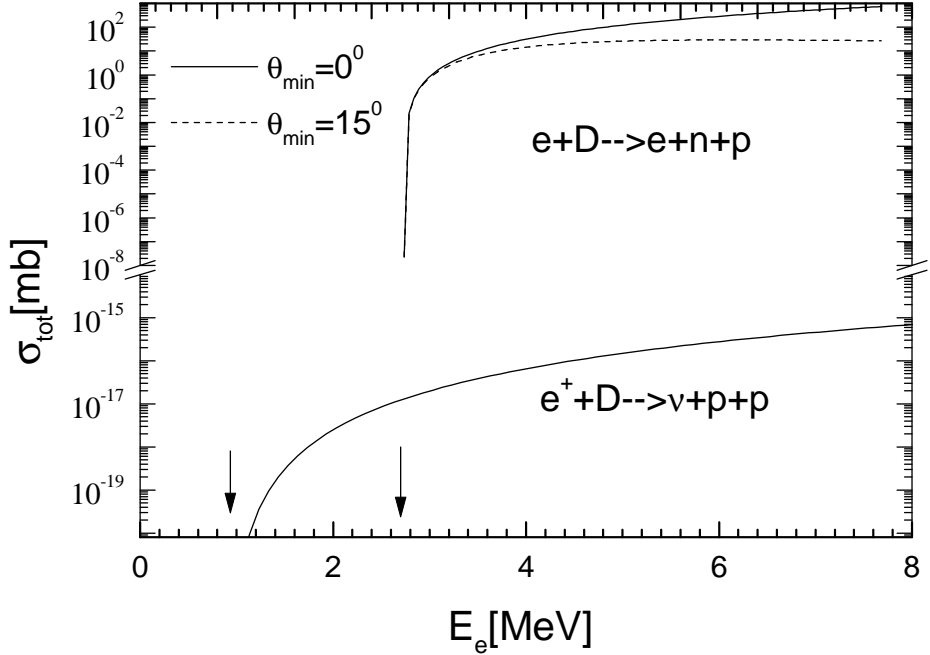}
\caption{The threshold behavior of the total cross section in electromagnetic
and weak disintegrations of the deuteron. The arrows depict the 
corresponding thresholds.
The dashed line displays the integrated $e D \to e n p$ cross section 
when the outgoing electron polar angle $\Theta$ is larger than 
$15^o$.
}
\label{fig2}
\end{figure}

In fig.~\ref{fig2} results of the numerical calculation of the 
total cross section
as a function of the  initial energy of the electron (positron) 
in the laboratory frame are displayed.  
Above the electromagnetic disintegration threshold there is 
probably no way to experimentally separate
the weak processes from the electromagnetic ones.
Below the electromagnetic and above the weak thresholds, 
the cross section of 
positron weak disintegration of the deuteron seems far too 
low to be measured under laboratory conditions. So one may conclude that 
at present an experimental investigation of the matrix element, relevant
for the  solar $pp$ fusion, is still impossible also when using the cross
channels. (This is in contrast to other reactions, cf.\ \cite{grosse}.) 
Only reliable theoretical
calculations  may be applied to estimate the nuclear reaction rates and to
interpret the corresponding  experimental neutrino  data.

{\bf III. Solar proton burning process:}
Let us now consider the process of the two-proton fusion
$
p_1\,p_2\,\to\,\nu e^+D
$
at low relative energies.
While this reaction looks rather similar to the processes in the previous section,  
here appears a peculiarity. Due to the relation
$2m_p > M_d+\mu $ 
(with $m_p=938.27231$ MeV, $\,\, M_d=1875.61339$ MeV,
deuteron binding energy $\varepsilon_D=2.22455$ MeV,  
and $ \mu=0.510999$ MeV)     
there is no energy threshold for the
$pp$ fusion  and, in principle, the fusion process  may
occur at arbitrarily small 
relative energies of the two protons. In this case, in the 
cross section (\ref{cs00}) two factors will play the major role:
(i) the two-proton flux ($\sim \sqrt{\lambda(s,m_p^2,m_p^2)}$) approaches 
rather rapidly zero and the computation of the cross section becomes 
problematic,
(ii) a strong Coloumb repulsion between protons makes the amplitude
${\cal M}$ sharply decreasing and compensates singularities appearing from
the vanishing flux. In order to avoid uncertainties connected to these 
circumstances
one usually separates the contribution of Coulomb repulsion by factorizing
the Coulomb part of the two-proton final state, which at low
relative energies may be taken as  
$
\psi_{pp}=
\displaystyle\frac{\sqrt{2\pi\eta}}{\rm e^{\pi\eta}-1}
\approx\sqrt{2\pi\eta}\,\,\rm e^{-\pi\eta},
$
where $\eta=\alpha/v$ is the Sommerfeld penetration parameter, 
$v$  denotes the relative
velocity of the protons, and  $T_{pp}=m_pv^2/4$ is the corresponding 
relative kinetic energy;
the exponent  is known  as the 
Gamov penetration factor \cite{adelb}.
Then one introduces the notion of the  
$S_{pp}$ factor, which is determined only by the nuclear part 
of the interaction and is connected  with the cross section by
$
\sigma\equiv\displaystyle\frac{S_{pp}(T_{pp}) }{T_{pp}} \rm e^{-2\pi\eta}.
$
Then the rate of $pp$ fusion reactions can be written in
the form\cite{bahcall,riska}
$
\langle \sigma\,v\rangle_{pp}
=1.3005\times 10^{-18} \left(2/T_6^2\right )^{1/3} 
f S_{pp}(T_{pp}){\rm e}^{-\tau}\,\, {\rm cm^3 s^{-1}};
$
here $T_6$ is the temperature in units of $10^6$ K, 
$\tau \sim 15- 40$ dominates
the temperature dependence of the reaction rate,
$f$ is a screening factor calculated first by Salpeter \cite{salpeter},
and $S_{pp}$ is in MeV b. The $S_{pp}$ factor is needed
at the most probable interaction energy $T^{(0)}_{pp}$, 
which is in a range of 5 keV $\cdots$ 30 keV. 
However, in most analyzes in the literature, the value of
$S_{pp}$ is quoted at vanishing energy, obtained , e.g.\ within
the low-energy effective theory  \cite{riska,panda,wise}. 
Hence, for the sake of extrapolation to the  finite value 
of $T^{(0)}_{pp}$, a good estimate
of the associated derivatives is also required.

We now apply the same formalism as in the previous section, i.e.\
the relativistic impulse approximation,   to estimate
the low-energy behavior of  $S_{pp}$. 
The effect of initial state interaction is accounted for only in the 
Coulomb part of
the interaction by adopting the corresponding part of the above 
two-proton wave function 
$\psi_{pp}$. A treatment of the nuclear part
of the proton wave function as plane waves means that in the  matrix element
all transitions are allowed. Nevertheless, at low energies only the 
lowest partial
waves  contribute to the proton wave function and one may expect that the total
cross section will be determined  by the axial part of the current operator 
(\ref{tokD}) and correspondingly by the super-allowed 
$0^+\to 0^+$ Gamov-Teller transition.

In the limit $T_{pp}\to 0$, apart from the phase space volume, 
the value of the deuteron wave 
function $\psi(p_0,|{\bf p}|)$ at vanishing values of its
arguments becomes important. 
As can be seen from eq.~(\ref{psis}),
at $|{\bf p}|\to 0,\,2E_p\to 2m$ the wave function $\psi(p_0,|{\bf p}|)$
displays a sharp maximum. Our solution of the BS equation~\cite{solution}
has been obtained by 
solving a system of coupled  integral equations with Gaussian meshes
with a minimum  value of  $|{\bf p}|= 0.310848493733 $ MeV,
so that an extrapolation
to $|{\bf p}|=0$ needs to be employed. This procedure 
causes some numerical uncertainty which 
we estimate  to be about 10\% in the  final results.

\begin{figure}
\vskip 0.5cm
\epsfbox{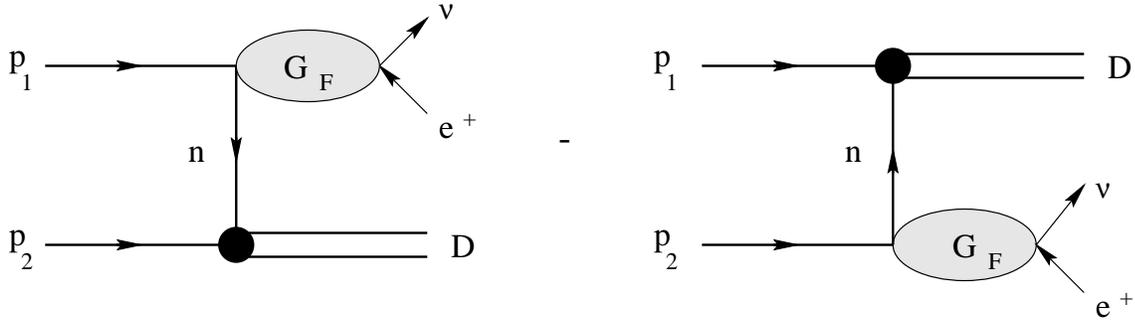}
\vskip 0.5cm
\caption{ Feynman diagrams contributing to the processes
$pp\to e^+\nu D$.}
\label{fig3}
\end{figure}

By observing that in the center of mass of colliding protons 
$2\sqrt{\lambda(s,m_p^2,m_p^2)}= 4E_1E_2\,v=sv$ 
and adopting the Gamow penetration factor, 
eq.~(\ref{cs00}) leads to
\begin{eqnarray}
&&
S_{pp}(T_{pp})=\frac{G_F^2}{8s}\frac{\pi m_p}{(2\pi)^5}
\int \frac{d{\bf P_d}}{E_d}\int \frac{d{\bf k_\nu'}}{{\cal E}_\nu'}
\left | {\cal M}_{pp\to e^+\nu D}\right |^2\delta(k_{e^+}^2-\mu^2),
\label{spp} 
\end{eqnarray}
where, due to the antisymmetrization of the initial two-proton state, 
the matrix element $ {\cal M}_{pp\to e^+\nu D}$  is now determined
by two Feynman diagrams, as depicted in fig.~\ref{fig3}, 
with a relative minus sign.
Correspondingly, the hadron tensor $W_{\mu\nu}$ consists of a direct and 
an interference term,
\begin{eqnarray}
\hspace*{-2.2cm}
W_{\mu\nu}^{(dir)} & = &
(p_2^2-m_p^2) {\rm Tr}
\left[\bar\Psi(p_1,p_n)
\Gamma_\mu (\hat p_2+m_p)\Gamma_\nu\Psi(p_1,p_n)(\hat p_2+m_p)\right] 
\nonumber\\
&& +
(p_1^2-m_p^2) {\rm Tr}
\left[\Psi(p_1,p_n)\Gamma_\nu (\hat p_1-m_p)\Gamma_\mu\bar\Psi(p_2,p_n)
(\hat p_1-m_p) \right],
\label{direct}\\
W_{\mu\nu}^{(int)} & = &
(p_2^2-m_p^2)(p_1^2-m_p^2) {\rm Tr}\left[
\bar\Psi(p_2,p_n)\Gamma_\nu\Psi(p_1,p_n)\Gamma_\mu
+
\bar\Psi(p_1,p_n)\Gamma_\mu\Psi(p_2,p_n)\Gamma_\nu
\right],\label{interf}
\end{eqnarray}
where $\Gamma_\mu$ is the vertex
corresponding to the effective current operator (\ref{tokD}).
The direct part of the hadron tensor has exactly the same 
form as for the  reactions $e^+D\to \nu pp$ (cf.\ eq.~(\ref{tr})), 
resulting in
an ``elementary'' cross section like that in  eq.~(\ref{elemw}). 
The interference term is different and  much more cumbersome, and
for the sake of brevity we do not display  it here. 

\begin{figure}
\vskip 0.5cm
\epsfxsize 4in
\centerline{\epsfbox{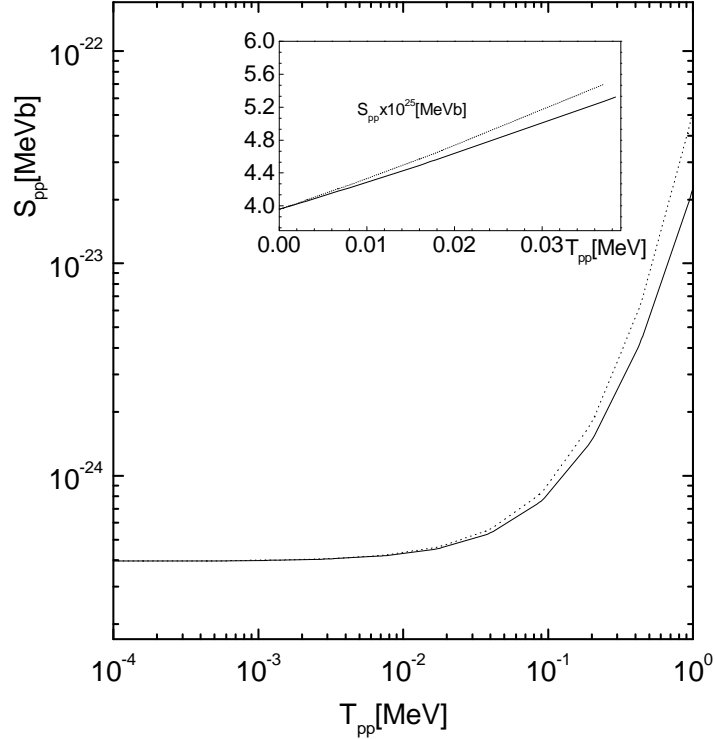}}
\caption{The $S_{pp}$ factor for the reaction $pp\to e^+\nu  D$
as a function of the relative kinetic 
energy of incident protons. The full line is the result of a calculation by
eq.~(\ref{spp}), the dotted line is the result of a computation 
with the deuteron
wave function taken as constant, $\Psi(p_0,|{\bf p}|)=\Psi(0,0)$. 
The inset shows $S_{pp}$ vs.\ $T_{pp}$ in  linear scales. 
}
\label{fig4}
\end{figure}

In fig.~\ref{fig4} the energy dependence of  $S_{pp}$  computed by 
eq.~(\ref{spp}) is displayed. 
The $S_{pp} $ factor, starting from 
$T_{pp} + 0$ up to
solar energies   $T_{pp}^0$,
displays a rather weak energy dependence. At very low kinetic energies
the value of $S_{pp}$ remains fairly constant, having a value
of $3.960\times 10^{-25}$ MeV b  in agreement with previous 
results \cite{adelb,kamion}. At larger  energies, however, 
the $S_{pp}$ factor
increases rather rapidly. The two curves in fig.~\ref{fig4} correspond to two
different methods of numerical calculations. In one case  
the deuteron wave function is
replaced by its value at zero transferred momenta. 
To some extent this  corresponds to calculations within low-energy effective 
potential models. The full line reflects results
of calculations without such an approximation. Only at relatively
high values of the kinetic energy, $T_{pp} >  20$ KeV, the effect of the 
deuteron wave function becomes sizeable. 

\begin{figure}
\epsfxsize 4in
\centerline{\epsfbox{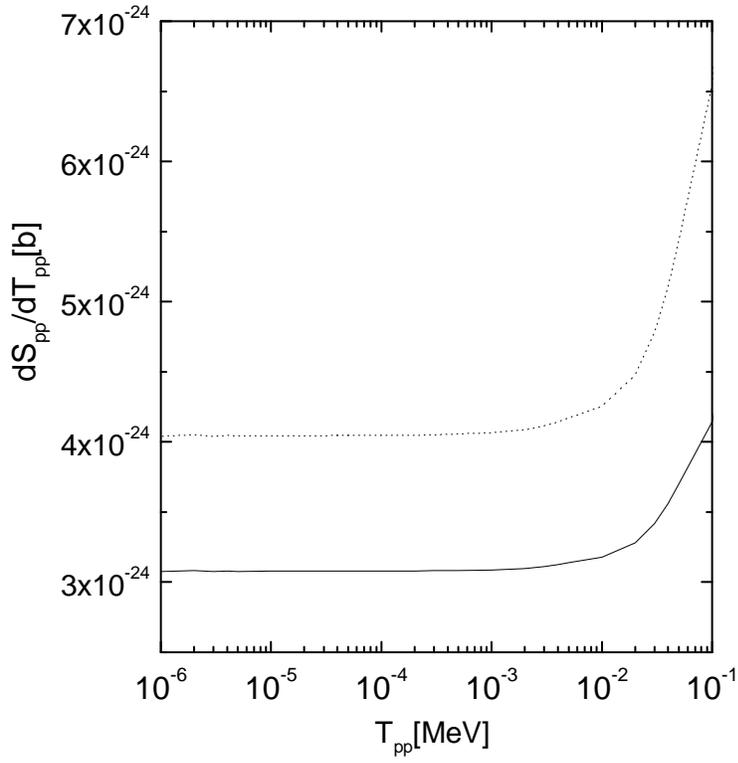}}
\caption{The first derivative of the low-energy nuclear reaction cross section
 factor $S_{pp}$.
Notation as in fig.~\protect\ref{fig4}}
\label{fig5}
\end{figure}

\begin{figure}
\epsfxsize 4.4in
\centerline{\epsfbox{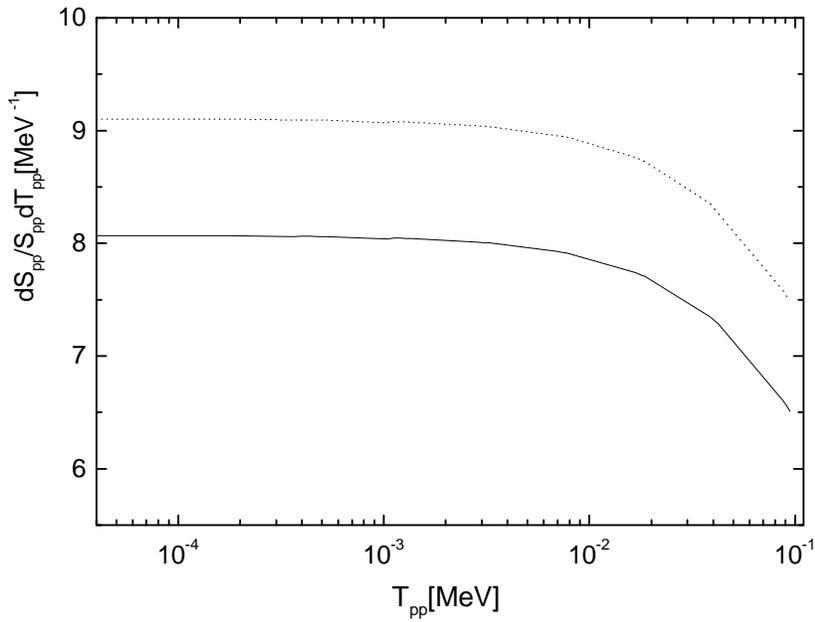}}
\caption{The logarithmic derivative of the low-energy nuclear 
reaction cross section factor $S_{pp}$.
Notation  as in fig.~\protect\ref{fig4}
}
\label{fig6}
\end{figure}

Within low-energy effective theories the associated derivatives of the 
$S_{pp}$ factor play  an important role.
The best estimates of the first
derivative seem to be  those quoted in ref.\cite{adelb}, 
$S'_{pp} = 4.48\times10^{-24}$ b, and
the logarithmic one $S'_{pp}/S = 11.2$ MeV$^{-1}$.
In figs.~\ref{fig5} and \ref{fig6} we display our estimates
for the energy dependence of these  derivatives. 
The derivatives are   computed at each value of $T_{pp}$
by interpolating  $S_{pp}$ with a quadratic interpolation formula.
From these pictures it is seen that within our approach
both the derivative and the 
logarithmic derivative are below the estimates in \cite{adelb} by 
$\sim 15-20 \%$.
However, at energies around  
$T_{pp}^{(0)}$  our results for the $S_{pp}$ factor and the results of
an interpolation with parameters from \cite{adelb} are rather close 
to each other. 

{\bf IV. Summary:}
We present a covariant model, based on the Bethe-Salpeter formalism,
to investigate the energy dependence of the 
matrix elements relevant to the solar proton-proton burning process.
Explicit covariant expressions for the cross section and the low-energy cross 
section factor $S_{pp}$ 
are presented. Our numerical estimates of $S_{pp}$ and its 
derivatives confirm previous results.

{\bf Acknowledgments:}
We thank H.-W. Barz, R. W\"unsch and  A.I. Titov  for useful discussions.
L.P.K. would like
to thank for the warm hospitality in the Research Center Rossendorf.
This work has been partially supported by
the Heisenberg-Landau JINR-FRG collaboration project, and by
BMBF grant 06DR829/1 and WTZ RUS-678-98.

\end{document}